\documentclass[twocolumn,amsmath,amssymb,floatfix,aps,prx,footinbib,superscriptaddress]{revtex4-1}

\usepackage{amsmath,amssymb,bm,graphicx,url,epsf,color}
\usepackage[english]{babel}
\usepackage{wasysym}
\usepackage{MnSymbol}
\usepackage{amsmath}
\usepackage{mathtools}
\usepackage{graphicx}
\usepackage{ulem}
\usepackage[colorinlistoftodos]{todonotes}
\usepackage[colorlinks=true, allcolors=blue]{hyperref}
\usepackage{subcaption} 
\usepackage[export]{adjustbox} 
\usepackage{cleveref}

\usepackage{graphicx}
\usepackage[export]{adjustbox}

\begin{document}

\title{Flying qubits Surfing on Plasmons}

\author{D.C. Glattli$^{\dagger\dagger}$}
\affiliation{SPEC, CEA, CNRS, Universit\'{e} Paris-Saclay, CEA Saclay, 91191 Gif sur Yvette  Cedex France}

\author{P. Roulleau$^{\dagger}$}
\affiliation{SPEC, CEA, CNRS, Universit\'{e} Paris-Saclay, CEA Saclay, 91191 Gif sur Yvette  Cedex France}

\date{\today}



\hspace{2cm}

\begin{abstract}
The rapid emergence of flying qubits in graphene and other low-dimensional conductors is pushing quantum electronics into an ultrafast regime where conventional transport theories no longer apply. In these systems, single-electron wave packets propagate coherently over micrometer scales while interacting with collective charge excitations on comparable time scales. Yet existing theoretical frameworks describe either fermionic single-particle dynamics or bosonic plasmonic modes, without reconciling the two. Here we introduce a unified theory of dynamical quantum transport that bridges this long-standing divide. Starting from a gauge-invariant scattering approach, we show how a time-dependent single-electron excitation self-consistently generates a propagating internal potential that behaves as a collective plasmonic mode. Electrons propagate at the Fermi velocity while simultaneously “surfing” on this self-induced plasmon wave, whose velocity is renormalized by Coulomb interactions and screening. This dynamical mean-field framework captures photon-assisted transport, charge relaxation, and edge magnetoplasmon dynamics within a single description and remains valid far beyond the low-frequency limit. By unifying single-electron and plasmonic pictures, our results provide a timely foundation for the interpretation and control of flying-qubit experiments in graphene at gigahertz–terahertz frequencies.
\end{abstract}

\maketitle

$^{\dagger}$ preden.roulleau@cea.fr

$^{\dagger\dagger}$ christian.glattli@cea.fr

\section{Introduction}

The ability to generate, manipulate, and detect ultrafast electronic excitations is a central goal of modern quantum nanoelectronics. In particular, the recent emergence of coherent electronic interferometers and flying-qubit architectures in graphene and other 2D electron platforms \cite{Assouline2023,Chakraborti2025,Ouacel2025,Shaju2025,Aluffi2023,Yamamoto2012} has opened access to regimes where single-electron wave packets propagate ballistically and coherently over micrometer distances on sub–nanosecond time scales. At the same time, rapid progress in ultrafast electronics has enabled the generation and control of voltage pulses spanning the gigahertz to terahertz frequency range. Together, these advances open access to a dynamical regime in which electronic motion, external driving, and Coulomb interactions occur on comparable time scales, creating new opportunities for time-resolved electron quantum optics, controlled emission of minimal excitations, and quantum interference experiments far beyond the adiabatic limit

These experimental advances place chiral quantum conductors well outside the domain of validity of conventional theories of ac quantum transport. For the experimental parameters (typical propagation lengths $L\sim1~\mu\mathrm{m}$ and single-particle Fermi (or drift) velocities $v_F\simeq5\times10^{4}~\mathrm{m/s}$), the characteristic traversal frequency $v_F/L$ lies in the tens of gigahertz range. As a result, GHz–THz excitations necessarily probe a regime in which the drive frequency $\Omega$ is comparable to, or larger than, the inverse propagation time. In this regime, the standard gauge-invariant scattering theory of ac transport developed by B\"uttiker and collaborators \cite{ButtikerThomasPretre1993,Buttiker1993,Pretre-Thomas-Buttiker1996}—remarkably successful at low frequencies—ceases to apply and must be extended. Its validity relies on the assumption that the internal electrostatic potential experienced by the electrons evolves slowly on the scale of their traversal time, leading to the constraint
\begin{equation}
\Omega \ll \frac{v_F}{L},
\end{equation}
which is explicitly violated in state-of-the-art graphene ps pulse experiments.

Beyond this low-frequency limit, a more subtle difficulty emerges: the coexistence of single-particle electronic excitations and bosonic collective charge dynamics. In one-dimensional conductors, Coulomb interactions reorganize charge transport into plasmonic modes, such as edge magnetoplasmons in quantum Hall systems. The pioneering work of X. G. Wen has long established the connection between interacting edge channels and chiral Luttinger liquids (LL)\cite{Wen1990}. The AC response of a chiral (LL) in the context of the Fractional Quantum Hall Effect has been computed in \cite{Crepieux2004} (see also \cite{Safi1999,Bena2007}). Traditional theoretical approaches tend to emphasize one description, namely bosonic versus fermionic, while neglecting the other. On the one hand, scattering theory provides a natural framework for fermionic single-electron excitations and photon-assisted transport but treats interactions in a largely static or local manner while enabling the description of two-path interference effects like the Aharonov-Bohm effect. On the other hand, bosonization and Luttinger-liquid theories capture collective modes elegantly but obscure the fate of individual fermionic excitations, which reappear only through tunneling operators while they do not offer a natural frame to describe two-path interference effects. A unified framework capable of treating single electrons and collective plasmons on equal footing is therefore highly desirable, especially in the ultrafast regime relevant to flying qubits.

In this work, we propose such a unified description of dynamical transport in chiral conductors. Our approach builds on and extends the self-consistent scattering theory developed by Pr\^etre, Thomas and B\"uttiker \cite{ButtikerThomasPretre1993,Buttiker1993,Pretre-Thomas-Buttiker1996} and extended to higher frequencies by Blanter, Hekking, and B\"uttiker (BHB) \cite{BlanterPRL1998,BlanterEPL1998}, which reconciles gauge invariance, current conservation, and dynamical screening. In these approaches, the self-consistent mean field internal potential is treated as a semiclassical variable. To better understand the physics at play, we will keep the choice of considering collective plasmonic modes classical. Further bosonization should be straightforward and is beyond the scope of this work. 

Using the paradigmatic example of an open mesoscopic capacitor formed by a chiral edge channel capacitively coupled to a gate, we derive the full ac response self-consistently. We show that the internal electrostatic potential generated by screened Coulomb interactions propagates as a collective edge magnetoplasmon with a velocity renormalized by both quantum and geometric capacitances. Remarkably, single electrons injected from the contacts propagate at the bare Fermi velocity while acquiring a dynamical phase imposed by this collective mode. The resulting picture is that of electrons ``surfing'' on their self-generated plasmon wave—a transparent physical interpretation that naturally reconciles fermionic and bosonic descriptions.

The paper is organized as follows. In Sec.~II, we recall the low-frequency regime \cite{Buttiker1993,Pretre-Thomas-Buttiker1996} in which the applied ac voltage varies slowly compared to the electron propagation time under the gate. In this limit, the internal potential remains uniform along the edge channel, and the scattering problem can be treated using the frozen scattering matrix  approximation. Starting from the time-dependent scattering formalism, we derive the ac current. In this low-frequency regime, the system reduces to an effective electrical circuit consisting of a quantum capacitance in series with the geometric capacitance and a universal charge relaxation resistance quantum. This section establishes the physical and conceptual reference point for a further extension to the high-frequency regime discussed later. In Sec.~III, we address the high-frequency regime in which the driving frequency becomes comparable to, or exceeds, the inverse electron traversal time. In this regime, the assumption of a spatially uniform internal potential breaks down and the dynamics of charge and current necessarily acquire a spatial structure along the edge channel. To capture these effects within the scattering framework, we introduce a discretized-gate geometry in which the screening gate is divided into a series of $N$ segments, each locally satisfying the frozen-scattering condition. This construction allows us to extend the validity of the theory to much higher frequencies while preserving gauge invariance and current conservation. Solving the resulting self-consistent problem, we derive the quantum admittance that incorporates finite propagation times and collective charge dynamics, thereby providing a unified description of high-frequency transport well beyond the adiabatic limit. In Sec.~IV, we develop a self-consistent scattering description that explicitly distinguishes between single-particle propagation and collective charge dynamics in a chiral interacting edge channel. This approach is the chiral version of the approach developed by Blanter, Hekking, and B\"uttiker (BHB) \cite{BlanterPRL1998,BlanterEPL1998} for a non-chiral one-dimensional wire with screened Coulomb interaction. Starting from the ballistic response of noninteracting electrons, we introduce Coulomb interaction through the polarization kernel and demonstrate how a dynamical internal electrostatic potential emerges from the electronic motion itself. By coupling this microscopic charge response to electrostatics, we show that the internal potential acquires its own dynamics and propagates as a collective edge magnetoplasmon with a velocity that differs from the bare Fermi velocity. This section provides a unified physical picture in which fermionic quasiparticles propagate ballistically while simultaneously generating and interacting with a collective plasmonic field, thereby reconciling single-electron transport, collective charge propagation, and dynamical screening within a single scattering-based framework. In Sec.~V, we introduce a complementary Floquet scattering formulation that describes single-electron dynamics in the presence of a self-consistently generated, propagating internal potential. Treating the internal electrostatic field as a collective edge-magnetoplasmon mode, we show how its propagation imprints a dynamical phase on fermionic wave functions while leaving their ballistic motion intact. This approach establishes a transparent correspondence between Floquet sidebands and plasmonic propagation. Importantly, it clarifies where and how photo-assisted electron–hole excitations are generated in an interacting chiral conductor \cite{Safi2019,Safi2022,Safi2010}. We demonstrate that interactions enter solely through collective phase retardation, leading to a shift in the effective timing of single- and two-particle interference without degrading fermionic coherence. Note that, as warned in the introduction, the plasmonic mode is treated as classical to make the physics more transparent. This is legitimate for an external AC drive of large amplitude. A further step, not done in the present work, would be to quantize the plasmonic field and introduce finite temperature effects. In this case the bosonic quantum fluctuations are expected to weaken the single particle fermionic coherence over some length scale \cite{LeHur2006}.
Together, Sections IV through VI present three approaches to reconciling the features of single-particle and collective mode excitations. Section VII demonstrates that this framework provides a unified interpretation of flying-qubit experiments, minimal excitations such as Levitons, and time-resolved interferometry, including Mach–Zehnder and Hong–Ou–Mandel geometries, in graphene-based quantum Hall devices. Section VIII is the conclusion.

\begin{figure}[t]
    \centering
    \includegraphics[width=\columnwidth]{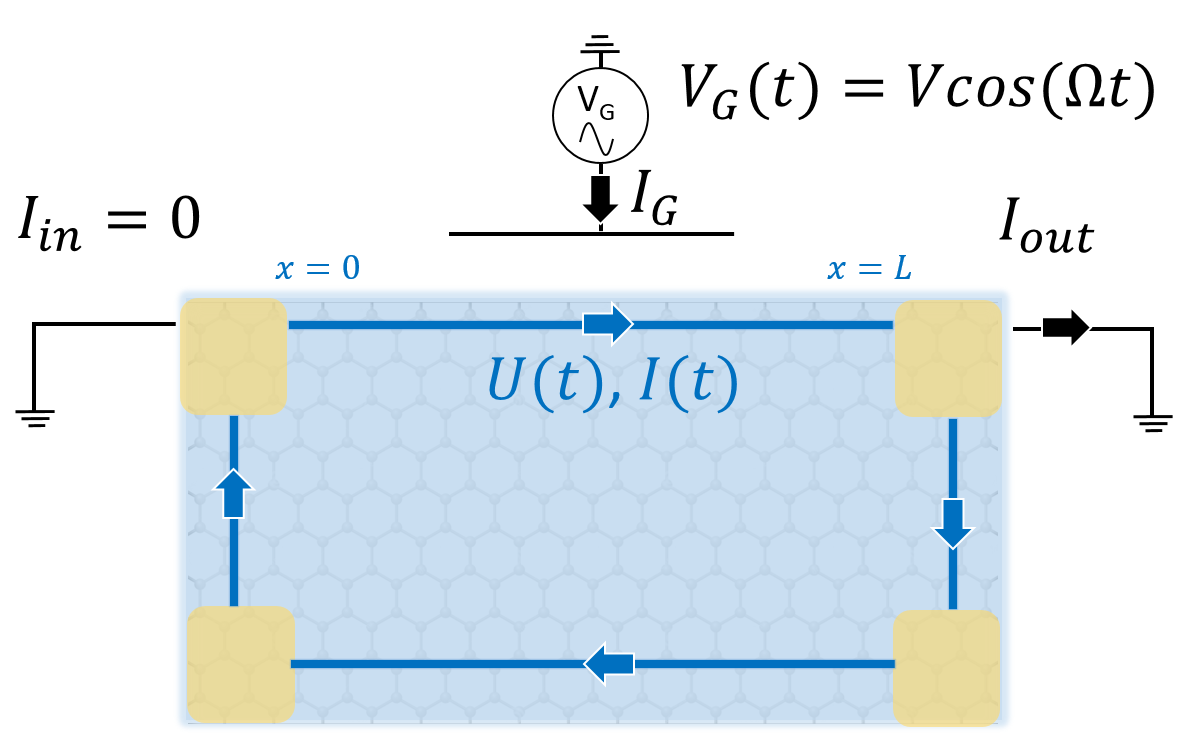}
    \caption{Open mesoscopic capacitor. The upper chiral edge channel (blue line), connecting ohmic contacts, is fully transmitted and propagates beneath a screening metallic gate. Together, the edge channel and the gate form a quantum capacitor. The gate is driven by a harmonic potential, and the resulting AC output current $I_{\mathrm{out}}$ is measured. The lower edge channel acts as a spectator and does not contribute to the AC current.}
    \label{fig:mesoscopic_capacitor_2}
\end{figure}

\section{Low Frequency regime}

\subsection*{\textit{A. Device geometry and physical regime}}

Figure~1 illustrates the system under consideration. A single chiral edge channel, connected to a grounded injecting contact, propagates ballistically from $x=0$ to $x=L$ beneath a metallic gate driven by a harmonic voltage
\begin{equation}
V_G(t)=\mathrm{Re}\!\left(V e^{i\Omega t}\right).
\end{equation}
The resulting time-dependent current is collected at a grounded output contact located at $x=L$,
\begin{equation}
I_{\mathrm{out}}(t)=I(t,x=L).
\end{equation}

Due to electrostatic screening by the gate, electrons experience a self-consistent internal potential
\begin{equation}
U(t)=\mathrm{Re}\!\left(U e^{i\Omega t}\right),
\end{equation}
which generally differs from the injecting potential. In the low-frequency (adiabatic) regime, we assume that $U(t)$ is spatially uniform along the gated region. This approximation is justified provided that the driving frequency satisfies
\begin{equation}
\Omega \ll \frac{v_F}{L},
\end{equation}
where $v_F$ is the single-particle Fermi (or drift) velocity of the chiral edge state.

In this regime, the displacement current induced by the gate,
\begin{equation}
I(t)\propto V_G(t)-U(t),
\end{equation}
is uniform along the edge, implying a spatially homogeneous charge accumulation beneath the gate. Current conservation then gives the total accumulated charge,
\begin{equation}
Q(t)=\int^{t}\!\mathrm{d}t'\,
\big[I(t',x=0)-I(t',x=L)\big],
\end{equation}
Here, the internal potential cannot develop any spatial structure, and the description excludes the spatially oscillating charge and potential profiles expected at higher frequencies. The extension beyond this adiabatic regime is addressed in Sec.~III-VI.

\subsection*{\textit{B. Floquet scattering computation of the ac current}}

The output current generated by the driven mesoscopic capacitor is evaluated using the time-dependent scattering approach introduced by B\"uttiker and collaborators and later extended to periodically driven systems~\cite{Buttiker1993,Splettstoesser2008}. This formulation is particularly well suited to the present problem, as it preserves gauge invariance and current conservation while allowing for a self-consistent treatment of displacement currents. Moreover, it naturally generalizes to geometries involving multiple gates and internal potentials, which will be required for the extension of the approach to the high-frequency regime discussed in section III.

For a periodic drive of angular frequency $\Omega$, the time-dependent current at the output contact can be expressed within the Floquet scattering framework~\cite{Brouwer1998,MoskaletsButtiker2002,Moskalets2011,Splettstoesser2008} as
\begin{equation}
\begin{split}
I_{out}(t)=\frac{e}{h}
\int_{-\infty}^{+\infty}\! d\varepsilon
\sum_{n=-\infty}^{+\infty}
\bigl[f(\varepsilon+n\hbar\Omega)-f(\varepsilon)\bigr] \\
\times \frac{1}{T}\int_{0}^{T}\! dt'\,
e^{in\Omega(t-t')}\,
S^*(t',\varepsilon)S(t,\varepsilon),
\end{split}
\label{eq:Io_floquet}
\end{equation}
where $T=2\pi/\Omega$ is the driving period, $f(\varepsilon)$ is the Fermi distribution in the injecting contact, and the integer $n$ labels photon-assisted processes involving the exchange of $n$ quanta of energy $\hbar\Omega$ with the drive.  Here $S(t,\varepsilon)$ is the scattering matrix \cite{Brouwer1998,MoskaletsButtiker2002} presented below.

For a single chiral edge channel with linearized dispersion relation $\varepsilon=\hbar v_F k$, the scattering amplitude factorizes as
\begin{equation}
S(t,\varepsilon)=e^{ik(\varepsilon)L}\,S(t),
\end{equation}
where the propagation phase $e^{ikL}$ accounts for free motion along the edge. The time-dependent part,
\begin{equation}
S(t)=
\exp\!\left[
\frac{i e}{\hbar}\int_{t-\tau}^{t}U(t')\,dt'
\right]
\equiv e^{i\Phi(t)},
\label{eq:phase}
\end{equation}
describes the dynamical phase acquired by the electron under the gate while experiencing the internal potential $U(t)$. Here $\tau=L/v_F$ is the propagation time through the gated region.

The energy integral in Eq.~\eqref{eq:Io_floquet} can be evaluated explicitly using the identity
\begin{equation}
\int_{-\infty}^{+\infty}\! d\varepsilon\,
\bigl[f(\varepsilon+n\hbar\Omega)-f(\varepsilon)\bigr]
= -n\hbar\Omega,
\end{equation}
which directly reflects the redistribution of electronic occupations induced by photon-assisted transitions. As a consequence, elastic processes ($n=0$) do not contribute to the current in the absence of a dc bias. The ac response therefore originates entirely from energy scattering channels ($n\neq0$), in which electrons are being absorbing or emitting energy quanta $\hbar\Omega$ from the drive.

The Floquet sum in Eq.~\eqref{eq:Io_floquet} can be recast into a compact and physically transparent form:
\begin{equation}
I_{out}(t)=\frac{e}{2\pi i}\,S(t)\,\frac{dS^*(t)}{dt}=(e/h)\frac{d\Phi}{dt}
\end{equation}
 The single particle propagation phase cancels out in the product
$S^*(t',\varepsilon)S(t,\varepsilon)$ appearing in Eq.~\eqref{eq:Io_floquet}, rendering the integrand independent of energy. As a result, the ac current is entirely governed by the time-dependent phase $\Phi(t)$ induced by the self-consistent internal potential.
\subsection*{\textit{C. Electrostatic screening and equivalent circuit}}

 Substituting the phase representation of the frozen scattering amplitude,
one obtains
\begin{equation}
I_{out}(t)=\frac{e^2}{h}\bigl[U(t)-U(t-\tau)\bigr],
\label{eq:Io_final}
\end{equation}
where $\tau=L/v_F$ is the propagation time under the gate.

Equation~\eqref{eq:Io_final} highlights a central physical result: the current probes the difference between the internal potential at the exit time $t$ and at the entrance time $t-\tau$. This nonlocality in time  manifests retardation effects intrinsic to fast ac driving.

Starting from the scattering result
Eq.~\eqref{eq:Io_final}, we first express the current in the frequency domain. For a harmonic time dependence at angular frequency $\Omega$, Eq.~\eqref{eq:Io_final} yields
\begin{equation}
I_{\mathrm{out}}(\Omega)
=
\frac{e^2}{h}\bigl(1-e^{-i\Omega\tau}\bigr)\,U(\Omega)
\equiv g_q(\Omega)\,U(\Omega),
\label{eq:Io_scattering}
\end{equation}
which defines the quantum admittance $g_q(\Omega)$ of the chiral channel.

Independently, current conservation requires that the output current equals the displacement current flowing through the gate capacitance $C$. In the frequency domain this gives
\begin{equation}
I_{\mathrm{out}}(\Omega)
=
i\Omega C\bigl[V(\Omega)-U(\Omega)\bigr].
\label{eq:Io_gate}
\end{equation}

Equating Eqs.~\eqref{eq:Io_scattering} and \eqref{eq:Io_gate} and solving for $I_{\mathrm{out}}(\Omega)$ leads to
\begin{equation}
I_{\mathrm{out}}(\Omega)
=
\frac{V(\Omega)}{1/g_q(\Omega)+1/(i\Omega C)},
\label{eq:Io_circuit}
\end{equation}
which reveals the driven mesoscopic capacitor as an effective series combination of the quantum impedance $1/g_q(\Omega)$ associated with electron propagation and the geometric capacitive impedance $1/(i\Omega C)$ associated with electrostatic screening.
In the low-frequency regime $\Omega\tau\ll1$, required for the validity of the frozen scattering approximation, the inverse admittance can be expanded as \cite{Pretre-Thomas-Buttiker1996}
\begin{equation}
\frac{1}{g_q(\Omega)}
\simeq \frac{h}{2e^2}+\frac{1}{i\Omega C_q},
\qquad
C_q=\frac{e^2}{h}\tau,
\end{equation}
recovering the well-known equivalent circuit of a quantum capacitance $C_q$ in series with the geometric capacitance $C$ and the universal charge relaxation resistance or B{\"u}ttiker's resistance:
\begin{equation}
R_q=\frac{h}{2e^2}.
\end{equation}
The factor $1/2$ does not originate from spin degeneracy, which is neglected here, but from the single-reservoir geometry of the open mesoscopic capacitor, corresponding to one half of a Sharvin resistance~\cite{Pretre-Thomas-Buttiker1996,Gabelli2006}.

\subsection*{\textit{D. Physical interpretation and frequency dependence}}

The time-dependent gate voltage imprints a dynamical phase $\Phi(t)$ on the chiral electron wave as it propagates beneath the gate. The output current is therefore proportional to the time derivative of this phase, reflecting the fact that the ac response is governed by phase modulation rather than by energy redistribution.

Because electrons require a finite time $\tau=L/v_F$ to traverse the gated region, they effectively sample the internal potential over a temporal window of duration $\tau$. The resulting current is proportional to the difference between the internal potential at the exit and entrance times, as expressed in Eq.~\eqref{eq:Io_final}. For a harmonic internal potential,
\begin{equation}
U(t)=U_\Omega\cos(\Omega t),
\end{equation}
the output current takes the explicit form
\begin{equation}
\begin{split}
I_{out}(t)
&=\frac{e^2}{h}U_\Omega\bigl[\cos(\Omega t)-\cos(\Omega(t-\tau))\bigr] \\
&=\frac{2e^2}{h}U_\Omega
\sin\!\left[\Omega\!\left(t-\tfrac{\tau}{2}\right)\right]
\sin\!\left(\tfrac{\Omega\tau}{2}\right).
\end{split}
\end{equation}
This expression clearly illustrates the role of retardation effects. In the low-frequency limit $\Omega\tau\ll1$, the response is suppressed, recovering the quasi-adiabatic regime in which electrons experience an effectively uniform potential during propagation. In contrast, when $\Omega\tau\sim1$, the finite traversal time leads to a strong and oscillatory ac response, signaling the breakdown of the conventional low-frequency description and motivating the spatially resolved treatment developed in the next sections.

\begin{figure}[t]
    \centering
    \includegraphics[width=\columnwidth]{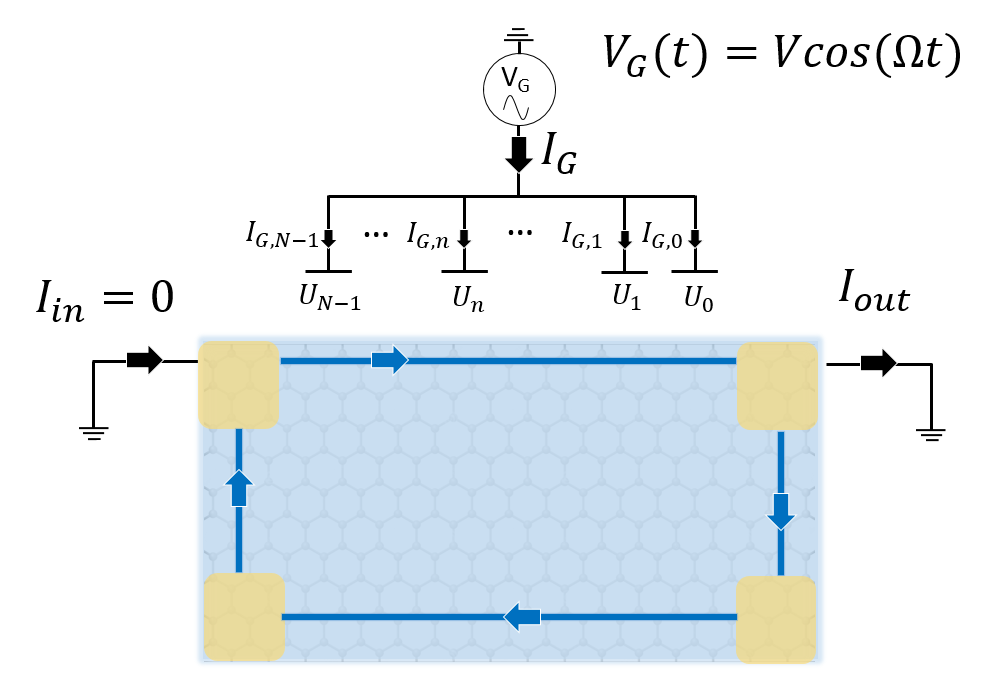}
    \caption{Going beyond the low-frequency limit. The metallic gate is partitioned into a series of $N$ shorter gates, all driven at the same potential $V\cos(\Omega t)$. Each segment induces an internal potential $U_n$ that is uniform over its length $L/N$. Within the frozen Floquet scattering approach, this spatial discretization extends the range of validity to $\Omega \ll Nv/L$, thereby circumventing the single-gate constraint $\Omega \ll v/L$.}

    \label{fig:mesoscopic_capacitor}
\end{figure}

\section{High-Frequency Regime and Discretized Capacitive Coupling}

\subsection*{\textit{A. Motivation and geometry}}

At higher frequencies, spatially varying displacement currents and charge densities necessarily develop. In this section, we show how introducing a spatially distributed capacitive coupling restores a controlled scattering description and allows one to access the high-frequency regime beyond the conventional adiabatic limit.
To extend the frozen scattering approach beyond the low-frequency limit, we consider the geometry shown in Fig.~\ref{fig:mesoscopic_capacitor}.  The screening gate is partitioned into $N$ identical sub-gates. All sub-gates are driven at the same potential $V(t)=\Re\!\left[V(\Omega)e^{i\Omega t}\right]$, so that the global electrical boundary conditions remain unchanged with respect to the previous model. Under the $n$th sub-gate ($n=0,\dots,N-1$), the edge channel experiences a local internal potential $U_n(t)$, which is assumed to be uniform over the short length $L/N$ all along the 1-D channel.

The displacement current flowing through sub-gate $n$ is then given by
\begin{equation}
I_{G,n}(\Omega)
= i\frac{C}{N}\Omega\bigl[V(\Omega)-U_n(\Omega)\bigr],
\end{equation}
where $C$ denotes the total geometric capacitance and $U_{n}(t)=\Re\!\left[U_{n}(\Omega)e^{i\Omega t}\right]$. This spatial discretization restores the possibility of position-dependent displacement currents and, consequently, of spatial variations of the edge-channel current amplitude, which are expected to develop at high driving frequencies.

\subsection*{\textit{B. Floquet Scattering formulation for a segmented gate}}

The output current is computed using the time-dependent scattering relation
\begin{equation}
I_{out}(t)=\frac{e}{2\pi}\frac{d\Phi(t)}{dt},
\end{equation}
where the accumulated phase $\Phi(t)$ now consists of additive contributions from each sub-gate. Explicitly,
\begin{equation}
\Phi(t)=\frac{e}{\hbar}\sum_{n=0}^{N-1}
\int_{t-(n+1)\theta}^{t-n\theta} U_n(t')\,dt',
\label{eq:phase_discrete}
\end{equation}
with $\theta=\tau/N$ the traversal time across a single segment of length $L/N$.

Taking the time derivative yields the output current in the time domain,
\begin{equation}
I_{out}(t)=\frac{e^2}{\hbar}\sum_{n=0}^{N-1}
\bigl[U_n(t-(n+1)\theta)-U_n(t-n\theta)\bigr],
\label{eq:Io_time_discrete}
\end{equation}
Defining $I_{out}(t)=\Re\!\left[I_{out}(\Omega)e^{i\Omega t}\right]$ , equation \ref{eq:Io_time_discrete} rewritten in the frequency domain becomes:
\begin{equation}
I_{out}(\Omega)=\frac{e^2}{\hbar}\bigl(1-e^{-i\Omega\theta}\bigr)
\sum_{n=0}^{N-1} U_n(\Omega)e^{-in\Omega\theta}.
\label{eq:Io_freq_discrete}
\end{equation}

\subsection*{\textit{C. Progressive internal potential wave ansatz}}

The frozen Floquet scattering approximation now requires $\Omega\theta\ll1$, ensuring that the internal potential remains uniform within each sub-gate. This condition extends the validity range of the approach by a factor $N$ compared to the single-gate geometry discussed in Sec.~II, while allowing for nontrivial spatial variations on the scale of the full gate length $L$. In this range of validity, we will make the ansatz that the internal potential propagates with the same single-particle chirality but with a velocity different from the single velocity . We thus assume that the discrete internal potentials can be expressed as $U_{n}(\Omega)=U_{0}e^{in\theta^{*}}$ where $\theta^{*}$ is to be determined self-consistently. Equation \ref{eq:Io_freq_discrete} now writes:
\begin{equation}
I_{out}(\Omega)=\frac{e^2}{\hbar}\bigl(1-e^{-i\Omega\theta}\bigr)
\sum_{n=0}^{N-1} U_{0}(\Omega)e^{-in\Omega(\theta-\theta^{*})}.
\label{eq:Io_freq_discrete_*}
\end{equation}
or, in the limit $\Omega\theta$ and $\Omega\theta^{*}$ $<<1$:
\begin{equation}
I_{out}(\Omega)=\frac{e^2}{\hbar}\frac{\theta}{\theta -\theta^{*}}
U_0(\Omega)(1-e^{-iN\Omega(\theta-\theta^{*}}).
\label{eq:Io_freq_discrete_*2}
\end{equation}

\subsection*{\textit{D. Self-consistent determination of the internal potentials}}

The next step is to relate the internal potentials $U_n$ to the drive voltage $V$. To this end, it is sufficient to determine $U_0$ by expressing that the displacement current in the subgate "0" is equal to the difference of the current living the corresponding section of the wire minus the current entering the wire. The subgate length being $L/N$, we can use equations \ref{eq:Io_final} and \ref{eq:Io_gate} valid for $\Omega \theta<<1$. This gives:

\begin{equation}
I_{G,0}(\Omega)= i\frac{C\Omega}{N}
\bigl[V(\Omega)-U_0(\Omega)\bigr]\simeq \frac{e^{2}}{h}i\Omega \theta.
\label{eq:self-consist}
\end{equation}
using $\frac{e^{2}}{h}\theta =  C_{q}/N$
we get:
\begin{equation}
U_{0}(\Omega)=\frac{V(\Omega)}{1+\frac{C_q}{C}}
\label{eq:self-consist-2}
\end{equation}
Inserting this result in equation \ref{eq:Io_freq_discrete_*2}:
\begin{equation}
I_{out}(\Omega)=\frac{e^2}{\hbar}\frac{\theta}{\theta -\theta^{*}}
\frac{V(\Omega)}{1+C_q/C}(1-e^{-iN\Omega(\theta-\theta^{*}}).
\label{eq:Io_freq_discrete_*3}
\end{equation}
To determine $\theta^{*}$, we take the limit $\Omega \Rightarrow 0$ for which $I_{out}(\Omega) \rightarrow i\Omega \frac{CC_q}{C+C_q}V(\Omega)$. This leads 
to $(\theta-\theta^{*})=\theta/(1+C_q/C)$ and the output current:

\begin{equation}
I_{out}(\Omega)=\frac{e^2}{\hbar}V(\Omega)(1-e^{-i\Omega\tau_{EMP}}). 
\label{eq:Io_freq_discrete_*4}
\end{equation}

We observe that the current no longer depends on the single-particle propagation time $\tau$ but from a new, shorter, time $\tau_{EMP}=L/v_{EMP}=L/v_F(1+C_q/C)$ where $v_{EMP}$ is the edge magneto-plasmon velocity:
\begin{equation}
v_{EMP}=v_F(1+C_q/C)
\label{eq:v_EMP}
\end{equation}
Equation \ref{eq:v_EMP} is consistent with the known edge magneto-plasmon velocity which was first implicitly derived in \cite{Volkov1998} taking the limit of screened Coulomb interaction. Equation \ref{eq:v_EMP} can be written as the sum of the single-particle velocity $v_F$ and the Coulomb induced magneto-plasmon velocity $v_{C}=v_{F}C_{q}/C=\frac{e^{2}}{h}\frac{1}{c}$, where $c=C/L$ is the geometrical capacitance per unit length. Early experiments in conventional GaAs/GaAlAs 2D systems measured only the second term as $v_F$ was negligible. However, in graphene the high $v_F$ was comparable to $v_C$ and has been recently measured \cite{Petkovic2013}.

In the next two following sections, we will present two more different approaches highlighting the interplay between single-particle and collective excitations. Section IV is an adaptation of the Blanter-Hekking-Büttiker (BHB) method for a non-chiral Luttinger liquid to the chiral case. Section V, is original and is obtained by solving the dynamics of a single particle surfing on the collective plasmon mode.  All three approaches exposed in sections (III, IV, and V) provide insight regarding the interplay between single-particle and collective excitations.

\section{From Single Electrons to Collective Plasmons}

In this section, we will solve the problem using the BHB approach~\cite{BlanterPRL1998} that was previously applied to a non-chiral 1D interacting wire, i.e., a Luttinger liquid.

\begin{figure}[t]
    \centering
    \includegraphics[width=\columnwidth]{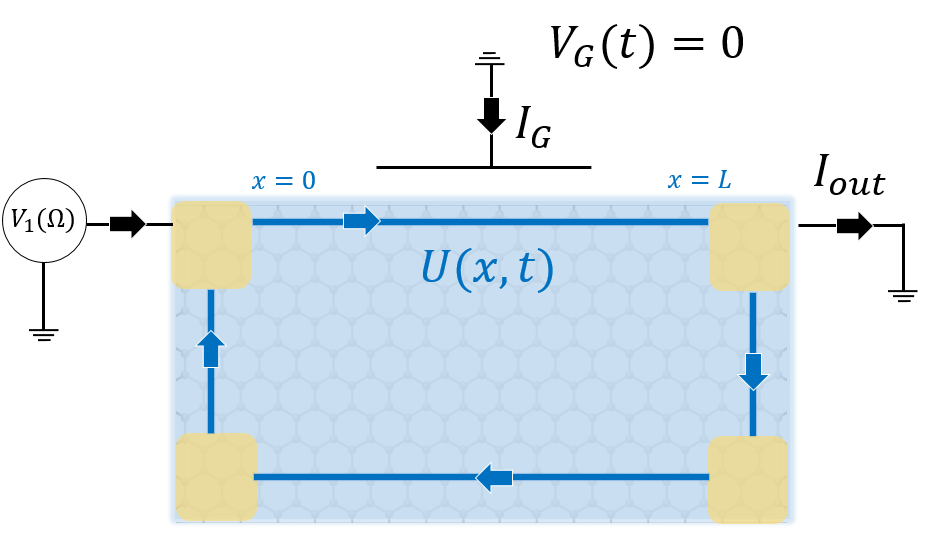}
    \caption{Schematic of the chiral gated edge channel. Electrons injected from left contact  propagate ballistically toward grounded contact. The gate-induced internal potential $U(x,t)$ mediates Coulomb interaction and supports collective edge magnetoplasmon modes.}
    \label{fig:plasmons}
\end{figure}

\subsection*{\textit{A. Geometry and driving protocol}}

Here we present the BHB approach. We consider electrons injected into a single chiral edge channel from left contact and propagating freely toward a contact to the ground, where they are absorbed. A nearby metallic gate screens the Coulomb interaction (Fig.~\ref{fig:plasmons}). A time-dependent voltage
\begin{equation}
V_1(t)=\Re\!\left[V_1(\Omega)e^{-i\Omega t}\right]
\end{equation}
is applied at left contact, located at $x=0$, while the gate and right contact are grounded. Our objective is to compute the current absorbed at the ground contact, located at $x=L$, as well as the local current
\begin{equation}
I(x,t)=\Re\!\left[I(\Omega,x)e^{-i\Omega t}\right]
\end{equation}
at an arbitrary position $x$ along the edge.

\subsection*{\textit{B. Ballistic response in the absence of interactions}}

We first recall the non-interacting limit, which serves as a useful reference. In the absence of Coulomb interaction, Floquet scattering theory yields a purely ballistic response. The AC current carried by non-interacting chiral electrons reads
\begin{equation}
I_0(\Omega,x)=\frac{e^2}{h}\,V_1(\Omega)\,e^{iq_F x},
\label{eq:I0}
\end{equation}
where $q_F=\Omega/v_F$ reflects the phase accumulated during propagation at the Fermi velocity. Using the continuity equation, the corresponding bare particle density is
\begin{equation}
\rho_0(\Omega,x)=\frac{1}{h v_F}\,e V_1(\Omega)\,e^{iq_F x}.
\label{eq:rho0}
\end{equation}
In this limit, electrons respond instantaneously to the external drive, and no internal electrostatic potential is generated. Transport is fully described by single-particle dynamics.

\subsection*{\textit{C. Coulomb interaction and emergence of an internal potential}}

When Coulomb interaction is present, the charge carried by propagating electrons is partially screened by the nearby metallic gate. This screening generates a self-consistent internal electrostatic potential $U(\Omega,x)$ along the channel, which feeds back on the electronic motion. As a result, the true particle density is no longer given by the bare response alone, but acquires an interaction-induced correction,
\begin{equation}
\rho(\Omega,x)
=
\rho_0(\Omega,x)
-
\int_0^L\!dx'\,
\Pi(x,x')\,eU(\Omega,x').
\label{eq:rho_int}
\end{equation}
The kernel $\Pi(x,x')$ is the polarization function introduced within the self-consistent scattering approach of Blanter, Hekking and Büttiker~\cite{BlanterPRL1998,BlanterEPL1998}. It plays a central role in the description of dynamical Coulomb effects: it encodes how an electrostatic perturbation created at position $x'$ and time $t'$ induces a density response at position $x$ and time $t$, taking into account both local compressibility and ballistic propagation along the channel. This formulation ensures gauge invariance and current conservation in the presence of time-dependent potentials, and provides a microscopic link between single-particle dynamics and collective charge response~\cite{ButtikerThomasPretre1993}.

\subsection*{\textit{D. Physical content of the polarization kernel}}

For a chiral channel, the polarization kernel takes the explicit form
\begin{equation}
\Pi(x,x')
=
\frac{1}{h v_F}
\Big[
\delta(x-x')
+
i q_F\,e^{iq_F(x-x')}\,\Theta(x-x')
\Big].
\label{eq:Pi}
\end{equation}
Each term has a transparent physical interpretation.

The local term $\delta(x-x')/(h v_F)$ describes the instantaneous compressibility of the electron gas. It accounts for the fact that a local change in the electrostatic potential immediately modifies the electron density at the same position, reflecting the finite density of states of the chiral edge channel.

The nonlocal term proportional to $e^{iq_F(x-x')}\Theta(x-x')$ captures the dynamical propagation of charge. It represents electrons that were perturbed by the internal potential at position $x'$ and subsequently propagated ballistically to position $x$ at velocity $v_F$. The Heaviside function enforces chirality: only electrons originating upstream ($x'<x$) can contribute to the density at $x$.

Taken together, these two contributions encode the coexistence of instantaneous local screening and nonlocal memory effects associated with ballistic propagation.

\subsection*{\textit{E. Collective Electrostatics Beyond Single-Particle Transport}}

Equation~\eqref{eq:rho_int} makes explicit that the internal electrostatic
potential $U(x,t)$ is not a passive background field imposed externally.
Rather, it is generated by the motion of the electrons themselves through
Coulomb interaction and, in turn, feeds back on their subsequent dynamics.
The charge density at a given position and time is therefore the result of a
self-consistent interplay between single-particle propagation and the induced
internal field.

A central point is that nothing in the polarization kernel
$\Pi(x,x')$ constrains the internal potential to propagate at the
single-particle drift velocity $v_F$. On the contrary, the nonlocal term in
Eq.~\eqref{eq:Pi}, which accounts for the ballistic propagation of density
perturbations along the chiral edge, implies that $U(x,t)$ obeys its own
dynamical equation. Once electrostatics is taken into account, this equation
admits propagating solutions that correspond to collective charge modes rather
than individual electrons.

These collective excitations are the edge magnetoplasmons. Their propagation
velocity $v_{\mathrm{EMP}}$ generally differs from $v_F$ and reflects both the
electronic compressibility of the channel and the strength of electrostatic
screening by the gate. Capturing this independent dynamics of $U(x,t)$ is
essential for any correct description of high-frequency transport, resonances,
and quantum interference. Any theory in which $U$ is artificially slaved to
single-particle motion inevitably suppresses collective effects and misses the
correct dynamical behavior.

To make this explicit, we now close the system of
Eqs.~\eqref{eq:rho0}–\eqref{eq:Pi} by relating the charge density to the
internal potential through a local capacitive relation,
\begin{equation}
U(\Omega,x)=\frac{e\,\rho(\Omega,x)}{c},
\label{eq:cap}
\end{equation}
where $c$ is the geometrical capacitance per unit length. This relation
expresses the electrostatic energy cost associated with accumulating charge
under the gate and provides the link between charge dynamics and the internal
potential.

Substituting Eq.~\eqref{eq:cap} into Eq.~\eqref{eq:rho_int} yields the
self-consistent integral equation
\begin{equation}
\begin{aligned}
\frac{c}{c_q}\,eU(\Omega,x)
={}& eV_1(\Omega)e^{iq_F x}
- eU(\Omega,x) \\
&- i q_F \int_0^x dx'\,
e^{iq_F (x-x')} eU(\Omega,x'),
\end{aligned}
\label{eq:U_integral}
\end{equation}
where $c_q=e^2/(h v_F)$ is the quantum capacitance per unit length. This
equation explicitly shows how the internal potential is driven by the injected
electrons, screened locally, and redistributed nonlocally by ballistic
propagation along the edge.

Following Ref.~\cite{BlanterPRL1998}, we seek a propagating-wave solution
of the form $U(\Omega,x)=A e^{iqx}$. Solving
Eq.~\eqref{eq:U_integral} then yields a renormalized wave vector
\begin{equation}
q=q_F \frac{c}{c+c_q},
\label{eq:q}
\end{equation}
reflecting the fact that interactions slow down the propagation of the
internal potential relative to the bare single-particle phase velocity.

The resulting internal potential and current read
\begin{align}
U(\Omega,x) &= \frac{c_q}{c+c_q}\,V_1(\Omega)e^{iqx},
\label{eq:Ufinal} \\
I(\Omega,x) &= \frac{c+c_q}{c_q}\frac{e^2}{h}U(\Omega,x)
= \frac{e^2}{h} V_1(\Omega)e^{iqx}.
\label{eq:Ifinal}
\end{align}
Equation~\eqref{eq:Ifinal} makes transparent how the current is ultimately
controlled by the internal potential: the current is proportional to the time
derivative of the collective phase accumulated under $U$, while interactions
redistribute the phase and charge between kinetic and electrostatic
contributions.

The physical picture that emerges is that of a composite excitation: the true
charge density $e\rho$, the internal potential $U$, and the total current $I$
propagate together as a collective mode with the edge magnetoplasmon velocity
\begin{equation}
v_{\mathrm{EMP}}
= v_F\!\left(1+\frac{c_q}{c}\right)
= \left(\frac{1}{c_q}+\frac{1}{c}\right)\frac{e^2}{h}.
\label{eq:vemp}
\end{equation}
This velocity is consistent with that derived from the discrete gate approach in Section IV, equation \ref{eq:v_EMP}. It continuously interpolates between the noninteracting limit
$v_{\mathrm{EMP}}\to v_F$ and the strongly screened regime ($c\rightarrow\infty$) where interaction vanishes. Crucially, it differs from the single-particle velocity
whenever a Coulomb interaction is present.

This result constitutes the central message of our work: dynamical transport
in chiral conductors cannot be understood solely in terms of single electrons
or purely bosonic modes. Instead, it requires a self-consistent description in
which fermionic propagation and collective electrostatics are treated on equal
footing, with the internal potential $U(x,t)$ emerging as an autonomous
dynamical field.
\subsection*{\textit{F. Discussion}}

Starting from ballistic single-particle
propagation at the Fermi velocity, and incorporating the local displacement
current generated by screened Coulomb interactions, the theory naturally yields
a self-consistent description in which both the internal potential and the
current propagate as collective plasmonic modes. The resulting dynamics are in
quantitative agreement with those obtained from Luttinger-liquid theory, while
remaining firmly rooted in a scattering-based, fermionic language.

A key strength of this approach is that it retains an explicit single-particle
description throughout. Unlike bosonization-based formulations, in which
fermionic degrees of freedom are entirely traded for collective fields, the BHB
framework makes transparent how collective charge dynamics emerge from the
underlying motion of individual electrons. In this sense, it offers a clear and
intuitive reconciliation between fermionic transport and collective bosonic
excitations. The main limitation of this approach is its inability to realistically model inter-channel tunneling processes, such as those induced by quantum point contacts. These processes require the bosonization of plasmonic modes, a task that is naturally handled within the Luttinger-liquid formalism.

The physical picture that emerges is that of electrons propagating ballistically
at the Fermi velocity while effectively ``surfing'' on a self-consistent internal
electrostatic potential that propagates at the faster edge magnetoplasmon
velocity. Transport thus involves two distinct but coupled dynamical processes:
the propagation of fermionic quasiparticles and the propagation of collective
charge fluctuations generated by Coulomb interaction.

\section{Floquet Scattering approach to a single-particle electronic wave Surfing on a Plasmonic wave}

In this section, we present a complementary formulation of dynamical transport based on a Floquet scattering approach at the single-particle level. This perspective provides another intuitive bridge between fermionic transport and collective plasmonic dynamics. The central physical picture is that of an electron emitted from left contact which propagates chirally along the edge at the bare Fermi velocity $v_F$, while interacting with a self-consistent internal electrostatic potential generated by collective charge motion. This internal potential is not static: It propagates along the edge as an edge magnetoplasmon with velocity $v_{\mathrm{emp}}$, reflecting the self-consistent dynamical screening by the gate of the interaction.

\subsection{Single-particle dynamics in a propagating internal potential}

We consider the time-dependent Schrödinger equation describing a right-moving electron in the presence of a space--time dependent electrostatic potential,
\begin{equation}
\left(
- i\hbar v_F \frac{\partial}{\partial x}
+ e U\!\left(t-\frac{x}{v_{\mathrm{emp}}}\right)
\right)\psi(x,t)
= i\hbar \frac{\partial}{\partial t}\psi(x,t).
\label{eq:schrodinger}
\end{equation}
The potential is assumed to take the form of a rigidly propagating wave, $U(t-x/v_{\mathrm{emp}})$, consistent with the collective nature of the edge magnetoplasmon derived in Sec.~III. This assumption encapsulates the fact that charge disturbances propagate coherently along the edge at the plasmon velocity rather than following individual electrons.

We seek solutions corresponding to an electron emitted from the reservoir at a well-defined energy $\varepsilon$. Owing to chirality and the absence of backscattering, the wave function naturally factorizes into a free-propagation part and a dynamical phase factor,
\begin{equation}
\psi_\varepsilon(x,t)
=
\exp\!\left[-\frac{i\varepsilon}{\hbar}\left(t-\frac{x}{v_F}\right)\right]
\exp\!\left[-i\varphi\!\left(t-\frac{x}{v_{\mathrm{emp}}}\right)\right].
\label{eq:ansatz}
\end{equation}
The first exponential describes ballistic propagation at the bare Fermi velocity $v_F$, which governs the phase coherence of fermionic excitations. The second exponential encodes the additional phase accumulated through interaction with the propagating internal potential.

Introducing the retarded time variable $u=t-x/v_{\mathrm{emp}}$ and denoting $\varphi'(u)=d\varphi/du$, substitution of Eq.~\eqref{eq:ansatz} into Eq.~\eqref{eq:schrodinger} yields
\begin{equation}
\varepsilon \psi_\varepsilon
+ \hbar \frac{v_F}{v_{\mathrm{emp}}}\varphi'(u)\psi_\varepsilon
+ eU(u)\psi_\varepsilon
=
\varepsilon \psi_\varepsilon
+ \hbar \varphi'(u)\psi_\varepsilon .
\end{equation}
The Schrödinger equation is therefore satisfied provided that the phase $\varphi(u)$ obeys
\begin{equation}
\varphi'(u)
=
\frac{e}{\hbar}\,
\frac{1}{1-v_F/v_{\mathrm{emp}}}\,
U(u).
\end{equation}
Integrating over $u$, we obtain the total phase accumulated by the electron along its trajectory,
\begin{equation}
\varphi(u)
=
\frac{e}{\hbar}\,
\frac{1}{1-v_F/v_{\mathrm{emp}}}
\int^u du'\, U(u').
\label{eq:phase}
\end{equation}

This expression highlights a key physical feature of the problem. Although the internal potential propagates collectively at the plasmon velocity $v_{\mathrm{emp}}$, the electron samples this potential while drifting at the slower Fermi velocity $v_F$. The relative motion between the electron and the plasmonic wave leads to the enhancement factor $1/(1-v_F/v_{\mathrm{emp}})$, which reflects the prolonged interaction time between the electron and the propagating electrostatic field. In this sense, the electron may be viewed as ``surfing'' on the plasmonic potential: its fermionic phase evolves at $v_F$, while its energy and timing acquire a nontrivial modulation determined by the collective charge dynamics. This separation between fermionic kinematics and plasmonic propagation lies at the heart of the coexistence of single-particle coherence and collective effects in chiral interacting conductors.

\subsection{Floquet scattering formulation of the current}

We now compute the time-dependent current using Floquet scattering theory, extending the standard formulation to the present situation in which the scattering phase depends on the retarded variable $u=t-x/v_{\mathrm{emp}}$ rather than on time alone. This generalization is essential to account for the fact that the internal potential experienced by the electron propagates collectively along the edge at the plasmon velocity $v_{\mathrm{emp}}$, while the electron itself drifts at the Fermi velocity $v_F$.

Within the Floquet framework, the current at position $x$ is given by~\cite{MoskaletsButtiker2002}
\begin{equation}
\begin{aligned}
I(x,t)
&=
\frac{e}{h}
\int d\varepsilon
\sum_{n=-\infty}^{+\infty}
\bigl[f(\varepsilon+n\hbar\Omega)-f(\varepsilon)\bigr] \\
&\quad \times
\frac{1}{T}
\int_0^T du'\,
e^{in\Omega(u-u')}
S^\ast(u',\varepsilon)S(u,\varepsilon).
\end{aligned}
\label{eq:floquet_current}
\end{equation}

where $T=2\pi/\Omega$ is the period of the drive and $f(\varepsilon)$ denotes the Fermi distribution in the injecting contact. This expression accounts for photon-assisted processes in which electrons scatter into different energies as a quantum superposition of states at an energy $\varepsilon+n\hbar \Omega$.

For a perfectly transmitting chiral edge channel, the scattering matrix reduces to a pure phase factor,
\begin{equation}
S(u,\varepsilon)
=
e^{ik(\varepsilon)x}\,e^{-i\varphi(u)},
\qquad
k(\varepsilon)=\frac{\varepsilon}{\hbar v_F}.
\end{equation}
The propagation phase $e^{ikx}$ reflects free ballistic motion at the Fermi velocity, while the dynamical phase $\varphi(u)$ encodes the interaction with the propagating internal potential. The energy-dependent propagation phase cancels out in the product
$S^\ast(u',\varepsilon)S(u,\varepsilon)$, leaving a result that depends only on the collective phase $\varphi$.

Using the phase accumulated along the electron trajectory, Eq.~\eqref{eq:phase}, the Floquet expression~\eqref{eq:floquet_current} can be evaluated explicitly. Remarkably, all details of the photon sidebands collapse into a simple and local relation between current and phase,
\begin{equation}
I(x,t)
=
\frac{e}{2\pi}\,\frac{d\varphi}{dt}
=
\frac{e^2}{h}\,
\gamma\,
U\!\left(t-\frac{x}{v_{\mathrm{emp}}}\right),
\label{eq:current_time}
\end{equation}
where the dimensionless factor
\begin{equation}
\gamma
=
\frac{1}{1-v_F/v_{\mathrm{emp}}}
=
\frac{c+c_q}{c_q}
\label{eq:gamma}
\end{equation}
naturally emerges.

Equation~\eqref{eq:current_time} has a clear physical interpretation. The current is proportional to the time derivative of the collective phase accumulated by the electron as it propagates under the gate. This phase, in turn, is generated by the internal potential associated with collective charge dynamics. The factor $\gamma$ plays the role of a Doppler-like enhancement: because the electron drifts at velocity $v_F$ while the internal potential propagates at the faster plasmon velocity $v_{\mathrm{emp}}$, the electron effectively samples the potential over an extended interaction time. As $v_{\mathrm{emp}}$ increases due to Coulomb interaction and screening, the same potential modulation produces a larger phase shift and hence a larger current response. The Floquet scattering formalism thus provides a transparent link between single-particle phase accumulation and plasmonic charge dynamics, making explicit how collective effects enter time-resolved transport without destroying fermionic coherence.


\subsection{\textit{Where are electron-hole pairs created in the Floquet picture?}}

The Floquet scattering formulation naturally raises a central physical question: 
where are photo-induced electron--hole pairs created in an interacting chiral conductor?
In the standard non-interacting picture, such excitations are often viewed as being generated locally at the driven contact, where the time-dependent voltage is applied. In the present interacting case, this intuition must be revised.

Here, the AC drive does not act directly on the electrons alone. This is the self-consistent internal potential $U(x,t)$ that propagates along the entire gated region as an edge magnetoplasmon which acts on electrons. As a result, photo-assisted processes occur all along the channel and are intrinsically tied to the collective dynamics of this internal field.\\

\textit{Floquet amplitudes in the presence of a propagating internal potential}\\

For a chiral channel, the scattering amplitude reduces to a phase acquired during propagation. In the presence of a time-dependent internal potential propagating at velocity $v_{\mathrm{emp}}$, the scattering matrix at position $x$ reads
\begin{equation}
S(t,\varepsilon)
=
e^{ik(\varepsilon)x}
e^{-i\varphi(t-x/v_{\mathrm{emp}})},
\label{eq:S_general}
\end{equation}
where the phase $\varphi$ is generated by the internal potential,
\begin{equation}
\varphi(u)
=
\frac{e}{\hbar}
\frac{1}{1-v_F/v_{\mathrm{emp}}}
\int^u du'\, U(u').
\end{equation}

We now assume a weak harmonic drive at frequency $\Omega$,
\begin{equation}
U(t)=U_\Omega \cos(\Omega t),
\end{equation}
and introduce the dimensionless drive strength
\begin{equation}
\alpha=\frac{eU_\Omega}{\hbar\Omega},
\end{equation}
with $\alpha\ll1$. The accumulated phase along the trajectory is: 
\begin{equation}
\varphi(t)
=
\alpha\!
\left[
\sin(\Omega t)
-
\sin\!\left(\Omega t-\frac{\Omega x}{v_{\mathrm{emp}}}\right)
\right].
\end{equation}

Expanding the scattering matrix to first order in $\alpha$, we obtain
\begin{equation}
S(t)\simeq e^{ik(\varepsilon)x}
\Bigl[
1+p_{1}e^{-i\Omega t}
+p_{-1}e^{i\Omega t}
+O(\alpha^2)
\Bigr],
\label{eq:S_floquet}
\end{equation}
with Floquet amplitudes
\begin{equation}
p_{\pm1}
=
J_1(\alpha)\,e^{\pm i\Omega x/v_{\mathrm{emp}}}.
\label{eq:p_pm}
\end{equation}

Equation \ref{eq:S_floquet} expresses that electrons are in a quantum superposition of different energies. The amplitude of probability $p_{\pm1}$ denote single-photon absorption and emission processes during propagation. Higher-order photon processes are disregarded in this weak-driving regime $\alpha<<1$.

\textit{Spatial structure of electron--hole pair creation.}

Equation~\eqref{eq:p_pm} reveals a key result: the Floquet probability amplitudes acquire a position-dependent phase proportional to $x/v_{\mathrm{emp}}$. This phase directly reflects the propagation of the internal potential at the edge magnetoplasmon velocity. Importantly, however, the corresponding absorption and emission probabilities,
\begin{equation}
|p_{\pm1}|^2 = J_1^2(\alpha),
\end{equation}
are independent of position and identical to those obtained in the non-interacting case.

This has a clear physical interpretation. While the phase of the photo-assisted processes is locked to the plasmonic propagation of the internal potential, the rate at which electron--hole pairs are generated is spatially uniform. Electron--hole pairs are therefore not created at a single localized point, such as the injecting contact, but are generated continuously along the entire length of the gated region.

\textit{Physical consequences: delocalized excitation and coherence.}

This delocalized creation mechanism is a direct consequence of Coulomb interaction. The AC drive excites a traveling internal electric field,
\begin{equation}
E(\Omega,x)=-\partial_x U(\Omega,x),
\end{equation}
which extends throughout the conductor and acts locally on the electrons as the plasmon passes by. Each infinitesimal segment of the edge thus acts as a weak source of electron--hole pairs, synchronized with the collective charge mode.
Despite this spatially distributed excitation process, fermionic coherence is fully preserved. The Floquet picture shows that the creation of electron--hole pairs is phase coherent and locked to the plasmon dynamics, rather than being associated with incoherent local scattering. This provides a transparent microscopic explanation for why interacting chiral conductors can sustain robust quantum interference even in the presence of strong Coulomb interaction.

In summary, the Floquet scattering approach reveals a qualitatively new picture of photo-assisted transport in interacting chiral systems: electron--hole pairs are generated coherently and continuously along the propagation path, with their phase evolution governed by the collective plasmonic dynamics of the internal potential.

\section{Levitons and Flying Qubits}

A particularly instructive illustration of the interplay between single-particle coherence and collective dynamics is provided by Leviton excitations \cite{Levitov1996,Keeling2006,Dubois2013} carrying a single electronic charge~$e$. A periodic train of Levitons can be generated by applying a periodic series of Lorentzian voltage pulses $V_1(t)$ with quantized Faraday flux $h/e$:
\begin{equation}
\frac{e}{\hbar}\int_t^{t+T} dt'\, V_1(t') = 2\pi,
\qquad
T=\frac{2\pi}{\Omega},
\label{eq:leviton_condition}
\end{equation}
which ensures the creation of a minimal excitation state consisting of exactly one additional electron above the Fermi sea and no accompanying electron--hole pairs [Levitov, Keeling].

From Section V, the internal potential $U(t)$ self-consistently generated  by the injected charge is related to the source voltage through:
\begin{equation}
V_1(t)=\gamma\,U(t),
\label{eq:leviton}
\end{equation}
with $\gamma=(c+c_q)/c_q$. As a result, a Lorentzian voltage pulse applied at the contact produces a Lorentzian internal potential propagating along the edge. Equation~\eqref{eq:leviton} then shows that the associated electronic wave packet preserves its Lorentzian shape while propagating rigidly at the edge magnetoplasmon velocity $v_{\mathrm{emp}}$.

The same conclusion follows directly from the current response,
\begin{equation}
I(x,t)
=
\frac{e}{2\pi}\frac{d\varphi}{dt}
=
\frac{e^2}{h}\,
V_1\!\left(t-\frac{x}{v_{\mathrm{emp}}}\right),
\label{eq:leviton_current}
\end{equation}
which describes a Lorentzian current pulse traveling at the collective velocity $v_{\mathrm{emp}}$. Importantly, this propagation velocity does not reflect the kinematics of a single electron, but rather the transport of charge by a collective plasmonic mode of the interacting edge channel.

At the level of Floquet scattering, the interaction manifests itself through position-dependent phase factors of the photo-absorption probability amplitudes $p_l$:
\begin{equation}
p_l(x)=p_l(0)\,e^{i l \Omega x / v_{\mathrm{emp}}},
\label{Eq.pl-Leviton}
\end{equation}
which encode the retardation associated with plasmon propagation. Here, we recall that only the probability amplitudes $p_l(0)$ with  $l\ge 0$ are non-zero for a Leviton inkected at $x=0$ by the contact. Crucially, we observe from equation \ref{Eq.pl-Leviton} that the structure of the Floquet amplitudes remains that of an ideal Leviton all along the channel with vanishing $p_{l<0}$, characterizing the complete absence of hole excitations. Electron--hole pairs are therefore not generated during propagation, despite the presence of interactions and collective dynamics.

This observation highlights a key physical consequence of interactions in a chiral conductor. While the Leviton retains its defining single-particle property as a minimal excitation state \cite{Keeling2006,Dubois2013}, its spatial and temporal evolution is governed entirely by the collective dynamics of the electron fluid. The electron itself propagates at the Fermi velocity $v_F$, but the charge it carries is transported by an edge magnetoplasmon propagating at $v_{\mathrm{emp}}$. In this sense, the Leviton can be viewed as a coherent fermionic excitation ``riding'' on a collective plasmonic wave. Measurements of AC current or charge alone therefore probe only plasmonic properties and are insensitive to the underlying fermionic phase coherence. Accessing genuine single-electron physics requires interferometric measurements sensitive to the fermionic phase, such as electronic Mach--Zehnder \cite{Ji2003,Roulleau2008,Jo2021,Chakraborti2025,Yamamoto2012} or Hong--Ou--Mandel interferometry \cite{Bocquillon2013,Dubois2013}.
\subsection{Flying qubits interferences}

\textit{The Aharonov Bohm Phase}\\

A recurrent concern in AC-driven chiral conductors is that experimentally accessible observables such as the current or the accumulated charge appear to be governed entirely by collective edge-magnetoplasmon (EMP) dynamics. This naturally raises the question of how genuine single-electron interference, as observed for instance in a Mach--Zehnder interferometer, can survive in the presence of strong Coulomb interactions. It is imperative to establish a clear distinction between the present work and that of Ref.\cite{Grenier2011}. In the latter, the quantum coherence of high-frequency collective electronic modes (plasmons) in electron quantum optics was discussed in the spirit of quantum optics for photons. The focal point of this study is the quantum interference of individual electronic particles in the context of self-induced neutral collective plasmonic modes.

Within the present scattering formulation, we establish a clear separation between two distinct phases: the fermionic phase, associated with the electronic dispersion and Fermi statistics, and the electrostatic phase, generated by the self-consistent internal potential. The fermionic phase propagates at the bare Fermi (or drift) velocity $v_F$, while the electrostatic phase propagates at the collective edge-magnetoplasmon velocity $v_{\mathrm{emp}}$.

Single-electron interference is therefore naturally described by coherently summing the quantum amplitudes associated with the different paths $k$ of lengths $L_k$ in the interferometer. For an electron emitted at energy $\varepsilon$, the wave amplitude along path $k$ reads
\begin{equation}
\begin{aligned}
\psi_{\varepsilon,k}(t)
={}&
\exp\!\left[-\frac{i\varepsilon}{\hbar}
\left(t-\frac{L_k}{v_F}\right)\right] \\
&\times
\exp\!\left[i\frac{e}{\hbar}
\int_{\mathcal{C}_k}\mathbf{A}\cdot d\boldsymbol{\ell}\right] \\
&\times
\exp\!\left[-i\,\varphi\!\left(
t-\frac{L_k}{v_{\mathrm{emp}}}\right)\right] .
\end{aligned}
\label{eq:psi_paths}
\end{equation}
where the second factor is the Aharonov--Bohm (AB) phase accumulated along path $\mathcal{C}_k$, and $\varphi$ denotes the dynamical phase imprinted by the propagating internal potential (the ``surfing'' phase). 

Equation~\eqref{eq:psi_paths} makes two essential points explicit. First, the AB phase depends only on the magnetic flux enclosed by the interferometer and is therefore completely insensitive to interactions and to the plasmonic dynamics. Coulomb interaction does not renormalize the magnetic phase, and the fundamental AB periodicity is preserved. Second, the dynamical phase associated with the internal potential enters as a common time-dependent modulation of the single-particle amplitude, propagating at $v_{\mathrm{emp}}$.

This separation of roles clarifies how chiral interferometers can simultaneously exhibit plasmon-dominated AC transport and robust single-electron interference. It also explains why interferometric measurements, rather than current measurements alone, are required to access fermionic coherence in interacting chiral conductors.
\\

\textit{The Hong-Ou-Mandel experiment}\\

\begin{figure}[h!]
    \centering
    \includegraphics[width=\columnwidth]{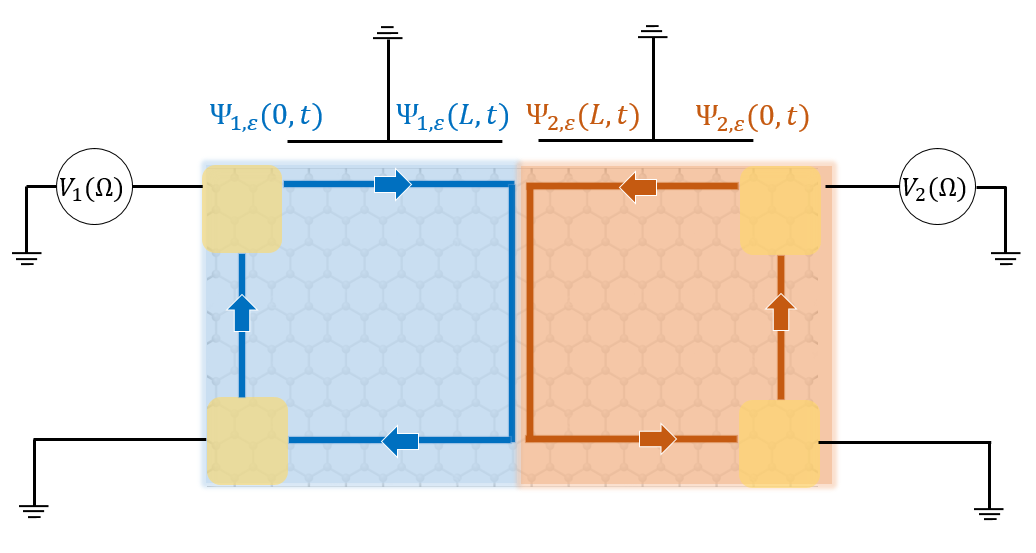}
    \caption{Hong--Ou--Mandel interference in a graphene $p$-$n$ junction. 
    Two time-dependent voltage pulses $V_1(t)$ and $V_2(t)$ are applied to opposite contacts, generating single-electron wave packets that propagate chirally toward a central $p$--$n$ junction acting as an electronic beam splitter. 
    The two excitations collide at the junction with a controllable time delay, leading to two-particle quantum interference.
    The resulting current correlations at the outputs provide a measure of fermionic Hong--Ou--Mandel interference.
   }
    \label{fig:HOM_graphene}
\end{figure}

Consider a Hong--Ou--Mandel (HOM) geometry in which two single-electron wave packets, emitted periodically and impinging on a QPC of transmission $D$ (reflection $1-D$), interfere with a relative delay $\tau$. Denote by $L_1$ and $L_2$ the propagation lengths from the sources to the QPC along the two incoming arms. The HOM noise can be written in the standard form (for a single channel and a periodic drive) as
\begin{equation}
S_I^{\rm HOM}(\tau;L_1,L_2)
=\frac{2e^2}{\pi}\,\Omega\,D(1-D)\,\bigl[1\pm g_2(\tau;L_1,L_2)\bigr],
\label{eq:SHOM_def}
\end{equation}
where the upper (lower) sign corresponds to bosonic (fermionic) exchange statistics, and $g_2$ is the (squared) modulus of the normalized first-order coherence overlap between the two incoming states at the QPC:
\begin{equation}
g_2(\tau;L_1,L_2)=\left|\frac{1}{T}\int_{-T/2}^{T/2}dt\;
\big\langle \hat\psi_2^\dagger(t+\tfrac{\tau}{2})\big|\hat\psi_1(t-\tfrac{\tau}{2})\big\rangle
\right|^2,
\label{eq:g2_def}
\end{equation}
where $T=2\pi/\Omega$ is the drive period and $\tau$ is the relative delay between the two sources at the QPC.
Within the Floquet scattering description, the effect of a periodic drive is encoded in the set of
photo-assisted (Floquet) amplitudes $\{p_l\}$, and the field operator at the QPC associated with a source
feeding an arm of length $L$ can be written (up to an overall dynamical phase that drops out from the
normalized overlap) as
\begin{equation}
\hat\psi_L(t)\propto \sum_{l=-\infty}^{+\infty} p_l(L)\,e^{-il\Omega t}\,\hat a_l,
\label{eq:psi_floquet}
\end{equation}
with $\hat a_l$ the annihilation operator of the incoming fermion from the corresponding reservoir.
The key interacting feature is that the Floquet amplitudes acquire a propagation phase locked to the
edge-magnetoplasmon velocity,
\begin{equation}
p_l(L)=p_{l}(0)\,e^{il\Omega L/v_{\rm emp}},
\label{eq:pl_phase}
\end{equation}
where $p_{l}(0)$ are the usual (position-independent) Floquet amplitudes of the noninteracting theory.

Using Eqs.~\eqref{eq:psi_floquet} and the orthogonality of different sidebands (the reservoirs are stationary),
one finds
\begin{align}
\Big\langle \hat\psi_{L_2}^\dagger(t+\tfrac{\tau}{2})\hat\psi_{L_1}(t-\tfrac{\tau}{2})\Big\rangle
&\propto \sum_{l} p_l^*(L_2)p_l(L_1)\,e^{il\Omega \tau}.
\label{eq:overlap_sum}
\end{align}
Therefore Eq.~\eqref{eq:g2_def} becomes
\begin{equation}
g_2(\tau;L_1,L_2)=\left|\sum_{l=-\infty}^{+\infty} p_l^*(L_2)p_l(L_1)\,e^{il\Omega \tau}\right|^2.
\label{eq:g2_floquet_sum}
\end{equation}

Using Eq.~\eqref{eq:pl_phase},
\begin{equation}
p_l^*(L_2)p_l(L_1)=|p_{l,0}|^2\,e^{-il\Omega(L_2-L_1)/v_{\rm emp}},
\end{equation}
so that Eq.~\eqref{eq:g2_floquet_sum} becomes
\begin{equation}
g_2(\tau;L_1,L_2)=\left|\sum_{l=-\infty}^{+\infty} |p_{l,0}|^2\;
\exp\!\left\{il\Omega\!\left(\tau-\frac{L_2-L_1}{v_{\rm emp}}\right)\right\}\right|^2.
\label{eq:g2_shifted}
\end{equation}
Thus, interactions do not change the weights $|p_{l,0}|^2$ of the sidebands; they only shift the
effective delay by the plasmonic flight-time difference $(L_2-L_1)/v_{\rm emp}$.\\

\textit{Example of a weak sinusoidal drive (single-photon regime).}\\

For a small harmonic drive $V(t)=V_{\rm ac}\cos(\Omega t)$, the Floquet spectrum is dominated by the
carrier and the first sidebands ($l=0,\pm1$). In that regime,
\begin{equation}
|p_{0,0}|^2\simeq 1-2\eta,\qquad |p_{\pm1,0}|^2\simeq \eta,\qquad |p_{|l|\ge2,0}|^2\ll \eta,
\label{eq:weak_drive_weights}
\end{equation}
with $\eta\propto (eV_{\rm ac}/2\hbar\Omega)^2$ (the precise prefactor is not needed for the phase dependence).
Keeping only $l=0,\pm1$ in Eq.~\eqref{eq:g2_shifted} gives
\begin{align}
\sum_l |p_{l,0}|^2 e^{il\Omega \Delta}
&\simeq |p_{0,0}|^2 + |p_{1,0}|^2 e^{i\Omega\Delta}+|p_{-1,0}|^2 e^{-i\Omega\Delta}\nonumber\\
&\simeq (1-2\eta)+\eta\left(e^{i\Omega\Delta}+e^{-i\Omega\Delta}\right)\nonumber\\
&= 1-2\eta+2\eta\cos(\Omega\Delta),
\label{eq:sum_cos}
\end{align}
where we defined the effective delay.
\begin{equation}
\Delta \equiv \tau-\frac{L_2-L_1}{v_{\rm emp}}.
\label{eq:Delta_def}
\end{equation}.

\textit{Discussion:}

Equation~\eqref{eq:sum_cos} shows that, within the ``electron surfing'' Floquet picture, Coulomb
interactions affect Hong--Ou--Mandel (HOM) interference only through a collective retardation
of the electronic phases. The relevant delay entering the overlap function is not the single-particle
time of flight $(L_2-L_1)/v_F$, but the plasmonic delay $(L_2-L_1)/v_{\rm emp}$ associated with the
propagation of the self-consistent internal potential. Equivalently, the HOM signal oscillates at the
drive frequency $\Omega$ as a function of the retarded delay $\Delta$ defined in
Eq.~\eqref{eq:Delta_def}, which is governed by edge-magnetoplasmon dynamics. This result makes explicit why AC observables in chiral conductors often appear ``purely plasmonic,''
while genuine single- and two-electron interference remains intact. Along each interacting arm, the
time-dependent drive is converted into a traveling internal potential that propagates at
$v_{\rm emp}$. This potential imprints a dynamical phase on the electron wave function and therefore
fixes when photo-assisted electron--hole components are generated along the path. As a result,
the effective timing of the wave packets reaching the beam splitter is shifted by the plasmonic
retardation $\Delta t_{\rm emp}$.

Crucially, however, the quantum point contact still mixes fermionic amplitudes locally and
instantaneously. The exchange process responsible for the HOM dip therefore remains a purely
fermionic effect, governed by the coherent superposition of single-electron amplitudes and by Fermi
statistics. This is reflected in the standard HOM structure of the noise, including the
statistics-dependent sign in Eq.~\eqref{eq:SHOM_def}. Interactions modify only the relative timing of
the incoming wave packets, not the interference mechanism itself. The present framework thus provides a transparent and unified description of two-electron
interference in interacting chiral conductors. One simply propagates each single-electron amplitude
according to Eq.~\eqref{eq:psi_paths}, accounting for the collective ``surfing'' phase associated with
$v_{\rm emp}$, and then combines these amplitudes at the scatterers. In this way, two-particle exchange
effects (as probed in HOM experiments) are naturally captured, while interactions enter solely
through the collective plasmonic dynamics of the internal potential.

\vspace{0.5cm}

\section{Conclusion}
In this work, we have developed a unified and physically transparent framework for dynamical quantum transport in interacting chiral conductors, tailored to the ultrafast regime now accessed by state-of-the-art flying-qubit experiments in graphene. By extending self-consistent scattering theory beyond its conventional low-frequency domain and explicitly disentangling single-particle propagation from collective charge dynamics, we have shown how fermionic quasiparticles can propagate ballistically while interacting with a self-generated, propagating internal electrostatic potential. This internal potential, which emerges naturally from Coulomb screening, propagates as a collective edge magnetoplasmon and imprints a dynamical phase on electronic wave functions without destroying fermionic coherence.

Beyond its immediate relevance to quantum Hall edge channels in graphene, the framework developed here is timely in view of the rapid expansion of the flying-qubit field. Novel chiral platforms are now emerging in twisted van der Waals heterostructures and in rhombohedral (ABC-stacked) graphene, where interaction-driven topological phases and the quantum anomalous Hall effect have recently been observed \cite{Lu2024,Lu2025}. The dissipationless chiral edge states hosted by these systems provide natural candidates for flying qubits, operating without external magnetic fields and with strong, tunable Coulomb interactions. Our results suggest that, in these platforms as well, ultrafast electron quantum optics will be governed by the same interplay between ballistic fermionic propagation and collective plasmonic dynamics elucidated here.

More broadly, this work bridges scattering theory, Floquet transport, and plasmonics within a single consistent description, providing practical tools and clear physical intuition for interpreting ultrafast experiments in chiral conductors. As flying-qubit architectures continue to evolve toward higher frequencies, shorter time scales, and more strongly correlated materials, we expect the concepts developed here to play a central role in the design, control, and interpretation of future electron quantum-optics experiments.\\

\begin{acknowledgments}
This work was funded by the ERC Starting Grant COHEGRAPH (Grant No.~679531) (P.R.), by the ANR project EQUBITFLY (P.R.), by the Horizon EIC 2022 Pathfinder Open project FLATS (P.R.), and by the Horizon EIC 2024 Pathfinder Open project ELEQUANT (P.R.).
\end{acknowledgments}

\end{document}